\documentclass[12pt,a4paper]{article}
\usepackage{psfrag}
\usepackage{epsfig}
\begin{document}
%
%
\title{A meron cluster solution for the sign problem of the two-dimensional
O(3) model}
\author{F. Brechtefeld\\
Inst. f\"ur Theoretische Physik III, Universit\"at Erlangen\\
Regionales Rechenzentrum Erlangen}
\date{\today}
\maketitle
\begin{abstract}
The two-dimensional O(3) model at a vacuum
angle $\theta=\pi$ is investigated. 
This model has a severe sign problem.
By a Wolff cluster algorithm an integer or half-integer
topological charge is assigned to each cluster. 
The meron clusters (clusters with half-integer topological charge)
are used to construct an improved estimator
for the correlation function of two spins at $\theta=\pi$. Only configurations
with 0 and 2 merons contribute to this correlation function.
 An algorithm, that generates configurations with only 0 and 2 merons,
is constructed and  numerical simulations at $\theta=\pi$ are performed.
 The numerical results indicate the presence of long range
correlations at $\theta=\pi$. 
\end{abstract}
\section{Introduction}
\label{intro}
The numerical simulation of many 
interesting physical systems, including in particular
 quantum chromodynamics (QCD)
 at non-zero chemical
potential or at non-zero theta vacuum angle, suffers from a complex
action problem or a sign problem (see \cite{Alf} and references
therein). The numerical simulation of these systems is
impossible with standard Monte Carlo methods.
In this article we investigate the two-dimensional O(3) model at a 
vacuum angle $\theta=\pi$. This model has a severe sign problem.
Unlike gauge theories, for which no reliable cluster algorithms are
available \cite{Jan,Car}, the O(3) model can be simulated with the
help of cluster algorithms and the meron cluster concept provides a
solution of the sign problem. The results obtained in our
studies are of interest for
the understanding of certain systems in statistical mechanics.
Haldane e.g.
 conjectured \cite{Hal} that integer and half-integer one-dimensional
antiferromagnetic quantum spin chains behave qualitatively differently. 
While integer spin chains have a mass gap, half-integer spin chains should be
gapless. In the large
$s$ limit ($2s+1$ being the number of possible spin orientations) there is a
mapping from the one-dimensional antiferromagnetic quantum spin chain on the
two-dimensional classical O(3) model with a $\theta$ vacuum term \cite{Aff}.
Integer spin chains correspond to $\theta=0$ and half-integer spin chains
correspond to $\theta=\pi$. The O(3) model at $\theta=0$ has a mass gap
in agreement with Haldane's conjecture. On the other hand, Haldane's
conjecture together with the (non-rigorous) mapping of spin chains on the O(3)
model implies that the mass gap disappears at $\theta=\pi$. This corresponds
to a phase transition governed by the vacuum angle $\theta$. 
\paragraph{}
The paper is organized as follows.
In section \ref{wca} the Wolff cluster
algorithm and the sign problem of the two-dimensional O(3) model at
$\theta=\pi$ are described. The meron clusters are defined. 
They are used for the construction of an improved estimator for
the correlation function of two spins at $\theta=\pi$ (section \ref{cfpi}).
 In order to calculate this correlation function 
a meron cluster algorithm is developed (section \ref{mca}).
 The improved estimator together
with the meron cluster algorithm
will be shown to solve the sign problem. 
The correlation function at $\theta=\pi$ is calculated
numerically. The results of these calculations are
 presented in section \ref{results}. 
\section{The Wolff cluster algorithm and the sign problem of the 
two-dimensional O(3) model}
\label{wca}
The lattice action of the two-dimensional O(3) model is given by
\begin{equation}
S= -\beta\sum_{\langle xy \rangle}e_x\cdot e_y, \qquad\beta>0.
\end{equation} 
The summation is over all pairs $\langle xy \rangle$ of
 nearest neighbor lattice sites. The three-component vectors of unit length
 are denoted by $e_x$ and called 
\emph{spins}. The action has a global O(3) rotation symmetry.
\par
In numerical simulations of this model one has to generate spin
configurations $\{e\}$ that are distributed with a probability given by
the statistical weight factor $\exp(-S(\{e\}))$. 
In the Wolff cluster algorithm
\cite{Wol1,Wol2} some spins are combined and 
then updated together in one step. For every pair of 
nearest neighbor lattice sites $x$ and $y$ there is a \emph{bond variable}
 $b_{xy}$ which only can take the values 0 and 1. If $b_{xy}=1$, one says
that $x$ and $y$ are connected by a bond. The bond variables define the
\emph{clusters}. A cluster is a set of lattice sites. Lattice sites 
connected  by bonds with each other belong to the same cluster. 
\par
The Wolff cluster algorithm consists of the following steps:
\begin{enumerate}
\item A vector $r$ of unit length is chosen whose direction (Wolff direction) 
is completely random.
\item To generate the clusters, nearest neighbor sites $x$ and $y$ are
connected with probability
\begin{equation} \label{bswskt}
p(e_x,e_y)= 1-\exp(\min[0,-2\beta(e_x \cdot r) (e_y \cdot r)]  ).
\end{equation}
(bond activation probability) by a bond.
 The probability is largest if
both spins $e_x$ and $e_y$ are aligned parallel or anti-parallel to the
Wolff direction $r$. It is zero if the projections of the spins on the
Wolff direction have different signs.
\item A cluster is flipped with probability $1/2$ independent of all the other
clusters. Cluster flip means that all the spins belonging to the cluster
are reflected on the plane perpendicular to the vector $r$ (Wolff plane)
\begin{equation} 
 e_x \rightarrow e^{\ast}_x = e_x - 2(e_x\cdot r)r.    
\end{equation}
$e^{\ast}_x$ denotes the reflected spin.
\end{enumerate}
It has been shown that this algorithm fulfills detailed balance
and is ergodic \cite{Wol1,Wol2}.
\paragraph{}
The spins of the two-dimensional O(3) model define a mapping from the
 lattice with periodic 
boundary conditions to the sphere $S^2$. These mappings fall  into different
classes characterized by an integer valued \emph{topological charge} $Q$.
For this characterization the spins must be interpolated between the lattice
points. The charge $Q$ counts, how many times the sphere is covered  by the 
mapping. The reason for
 the existence of the integer valued topological charge $Q$ is
that the homotopy group of continuous mappings  from the torus $T^2$
(the lattice) to the sphere $S^2$ is isomorphic to the group of integers.
\par
We consider the two-dimensional O(3) model with an additional topological
term in its action
\begin{equation} 
S \rightarrow S - {\rm i}\theta Q.
\end{equation} 
with $\theta$ real.
A complex probability distribution $\exp(-S)$ does not make sense.
 For the calculation
of expectation values the additional factor
 $\exp(\rm{i}\theta Q)$ must be included in the observable 
\begin{equation} \label{vzproblem}
\langle F \rangle \rightarrow \langle \exp({\rm i}\theta Q) F \rangle
\stackrel{\theta=\pi}{=}\langle (-1)^Q F \rangle.
\end{equation}
For the special case $\theta = \pi $ the additional factor becomes
$ (-1)^Q$. The factor $ (-1)^Q$ causes a \emph{sign problem}.
In a Monte Carlo calculation the alternating sign  $(-1)^Q$
 leads to cancellations
between the contributions from different configurations to the expectation
value. The numerical calculation of the expectation value
 is difficult because it has a large variance if $F$
is one of the usual physical observables.
\paragraph{}
Using a Wolff cluster algorithm the flip of a cluster changes the topological
charge from $Q$ to $Q^{\ast}$. We can assign to every cluster a 
\emph{cluster charge} \cite{Bie}
\begin{equation} \label{def_cl_ladung} 
q = \frac{Q-Q^{\ast}}{2}
\end{equation}
that is either integer or half-integer. Clusters with half-integer charge
are called \emph{meron clusters}. Note that the flip of a meron cluster
changes the sign $(-1)^{Q^\ast}=(-1)^{-2q}(-1)^Q=(-1)(-1)^Q$.
 This property of the meron clusters is exploited for
the construction of improved estimators at $\theta=\pi$ (see next
section).
\section{An improved estimator for the correlation function}
\label{cfpi}
The numerical calculation of the correlation function
$\langle (-1)^Q e_x \cdot e_y \rangle$
 of two spins at 
 $\theta=\pi$ suffers from large statistical errors because of
the sign problem. However the meron clusters can be used to construct an
improved estimator (see eq. (\ref{korrfkt_pi})) for the correlation function.
This construction is based on a proposal by U.-J. Wiese.
\par
Because of the O(3) rotation symmetry it is sufficient to
consider the product of the parallel components of the spins (parallel to
the Wolff direction). There are two possibilities: Either  the lattice sites
$x$ and $y$ belong to the same cluster ${\cal C}$ or they belong to
different clusters
\begin{eqnarray}
\lefteqn{\langle (-1)^Q e_x^{\parallel} e_y^{\parallel} \rangle= } \nonumber \\
& & \langle (-1)^Q e_x^{\parallel} e_y^{\parallel}
         \Theta({\cal C}_x={\cal C}_y) \rangle +
\langle (-1)^Q e_x^{\parallel} e_y^{\parallel}
         \Theta({\cal C}_x\not={\cal C}_y) \rangle .
\end{eqnarray}  
The symbol $\Theta(\quad)$ is 1 or 0,
if the statement in brackets is true or false respectively. 
${\cal C}_x$ denotes the cluster that contains the site $x$.
The flip of a meron cluster changes the sign $(-1)^Q$. Therefore when 
averaging over cluster flips the first term only gets a
non-vanishing contribution if the cluster configuration contains no merons. If
$x$ is contained in a meron cluster ${\cal M}_x$ the flip of the meron cluster
changes $(-1)^Q$ and $ e_x^{\parallel}$ simultaneously. Therefore when 
averaging over cluster flips the second term only gets non-canceling
contributions if $x$ and $y$ are both in meron clusters
 and there are no other 
meron clusters
\begin{eqnarray} \label{korr1_pi}
\lefteqn{\langle (-1)^Q e_x^{\parallel} e_y^{\parallel} \rangle= }  \\
& & \langle (-1)^Q e_x^{\parallel} e_y^{\parallel}
         \Theta({\cal C}_x={\cal C}_y)\Theta(N_{\cal M}=0) \rangle +
\langle (-1)^Q e_x^{\parallel} e_y^{\parallel}
         \Theta({\cal M}_x\not={\cal M}_y)\Theta(N_{\cal M}=2) \rangle
 \nonumber .
\end{eqnarray}
The factor $(-1)^Q$ can be removed in both terms on the right hand side.
 Every cluster configuration can be thought of as being
 generated by cluster flips from a configuration in which all spins 
have been on the same side of the Wolff plane. For this configuration
 $(-1)^Q=1$, because $Q=0$, and
$ e_x^{\parallel} e_y^{\parallel} > 0 $. The flip of a non-meron cluster
does not change $(-1)^Q$ and $ e_x^{\parallel} e_y^{\parallel}$. The 
flip of a meron cluster simultaneously changes $(-1)^Q$ and the sign of
 $ e_x^{\parallel} e_y^{\parallel}$ in the case
${\cal M}_x\not={\cal M}_y$. Therefore
\begin{eqnarray} \label{korrfkt_pi}
\lefteqn{\langle (-1)^Q e_x^{\parallel} e_y^{\parallel} \rangle= }  \\
& & 
\underbrace{\langle |e_x^{\parallel} e_y^{\parallel}|
         \Theta({\cal C}_x={\cal C}_y)\Theta(N_{\cal M}=0) 
\rangle}_{=g_{\mathrm{0M}}} +
\underbrace{\langle |e_x^{\parallel} e_y^{\parallel}|
         \Theta({\cal M}_x\not={\cal M}_y)\Theta(N_{\cal M}=2) 
\rangle}_{=g_{\mathrm{2M}}}
 \nonumber .
\end{eqnarray}
With equation (\ref{korrfkt_pi}) the sign problem has turned into the
problem of finding cluster configurations with only 0 or 2 merons.
 Usually cluster configurations contain many merons
and configurations with only 0 or 2 merons are quite rare. Therefore one 
aims at constructing an algorithm that generates only configurations with 0
and 2 merons. Such an algorithm is described in
the next section.
The improved estimator on the right hand side of equation (\ref{korrfkt_pi})
has no sign problem and therefore enables reliable numerical calculations.
\par
The first term $g_{\mathrm{0M}}$ only gets  non-vanishing contributions from 
cluster configurations in the zero-meron sector while
the second term  $g_{\mathrm{2M}}$
only gets non-vanishing contributions from the two-meron sector.
In the term $g_{\mathrm{0M}}$ the sites $x$ and $y$ must
 be in the same cluster.
Therefore, if the distance $r=r(x,y)$
between the lattice sites $x$ and $y$
 is much larger than the cluster size,
there is no contribution to $g_{\mathrm{0M}}(r)$.
The decay of
$g_{\mathrm{0M}}(r)$ is known to be mainly governed by the cluster 
size distribution
which is exponential \cite{Wol2}.
On the other hand in the term
$g_{\mathrm{2M}}$ the sites $x$ and $y$ must be in two different meron
clusters. The two meron clusters can be far apart. Then 
$g_{\mathrm{2M}}(r)$ gets non-vanishing contributions at
large distances $r$ (large compared to the cluster size).
 In this way long range correlations can arise.
\par
\begin{figure}
\noindent
\begin{minipage}[b]{.9\linewidth}
\psfrag{theta=0}{\tiny $\theta=0$} 
\psfrag{theta=pi}{\tiny $\theta=\pi$}
\psfrag{r}{$r$}
\centering\epsfig{file=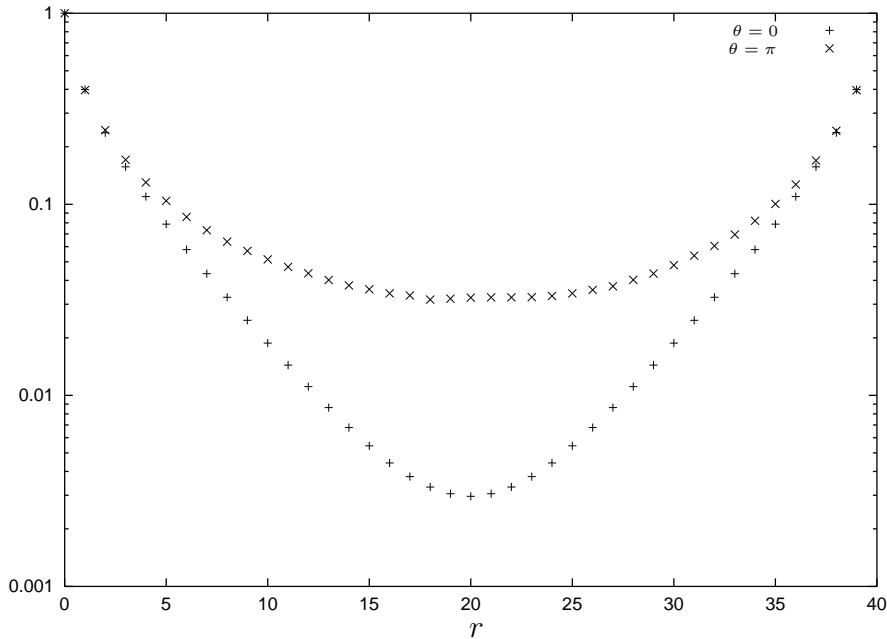,width=\linewidth}
\caption{\small \label{vergleich1} Correlation function at $\theta=0$ and
  at $\theta=\pi$. $L\times L = 40 \times 40$, $\beta=0.1$.}
\end{minipage}
\end{figure}

In figure \ref{vergleich1} the correlation functions
at $\theta=0$ and at $\theta=\pi$ are compared. 
This figure shows the slower decay of
correlation function at $\theta=\pi$.
\section{A meron cluster algorithm} 
\label{mca}
 Assume that we have a cluster configuration
with at most 2 merons. 
In the Wolff cluster algorithm the Wolff direction,
used in the definition of the clusters, is not fixed.
 After a change of the Wolff direction,
however, all bonds become invalid and have to be calculated anew. The new 
cluster configuration likely contains more than 2 merons.
\par
Thus, a gradual update (that is an update which changes only a few bonds in
each step)
 of the allowed cluster configuration is only
possible if the Wolff direction is fixed. In this case the spins
are only reflected on a single plane (perpendicular to the Wolff direction).
These reflections do not change the spin components perpendicular to the
Wolff direction.
Therefore the resulting algorithm is not ergodic. To make it ergodic we must 
combine it with another update step effecting the perpendicular spin
components. In this additional step only a few
spins and bonds should be updated because we want to make only small changes
on the allowed cluster configuration.
\paragraph{}
For the numerical calculations
we use an algorithm with two Wolff directions. Wolff direction 1 is fixed
during the whole simulation. The clusters and their charges are defined
with respect to this Wolff direction 1. The cluster flips on the corresponding
 Wolff plane are not ergodic. To establish ergodicity we use
a single cluster algorithm with a Wolff direction 2 that is chosen 
randomly anew for every update step. In one update step a single cluster is
calculated and flipped with respect to the Wolff direction 2. After the single
cluster flip the bonds in configuration 1 are calculated anew and the meron
number constraint for configuration 1 is checked. If the meron number
constraint is violated the single cluster flip is undone and the old bond
configuration 1 is restored (accept reject step). 
\par
After the single cluster flip only those bonds in configuration 1 are
calculated anew whose corresponding spins got changed. Thereby the change of
the configuration of spins and bonds induced by the update step is restricted
to a relatively small part of the lattice and the acceptance ratio for
the update step becomes independent of the lattice size.
\par
With the introduction of a second Wolff direction, the obvious problems with
ergodicity are removed. There remains however still the posibility that 
independent of the details of the algorithm non-ergodic behaviour could arise
if certain components of the zero- and two-meron sector are not accessible 
to our algorithm. It could be that configurations  in the  zero- and 
two-meron sector exist which could be reached only from sectors with more 
than two merons. This would invalidate any approach which deals with selected
meron sectors only. We have not been able to rule out this possibility 
analytically. However we have not found any dependencies on the initial
configuration in our results for the correlation function. 
\section{Numerical results}
\label{results}
\begin{figure}
\noindent
\begin{minipage}[b]{.9\linewidth}
\psfrag{fitfunh}{\tiny $h(r)$} 
\psfrag{2mercorrf}{\tiny $g_{\mathrm{2M,p}}(r)$}  
\psfrag{fullcorrf}{\tiny $g_{\mathrm{p}}(r)$}
\psfrag{r}{$r$}
\centering\epsfig{file=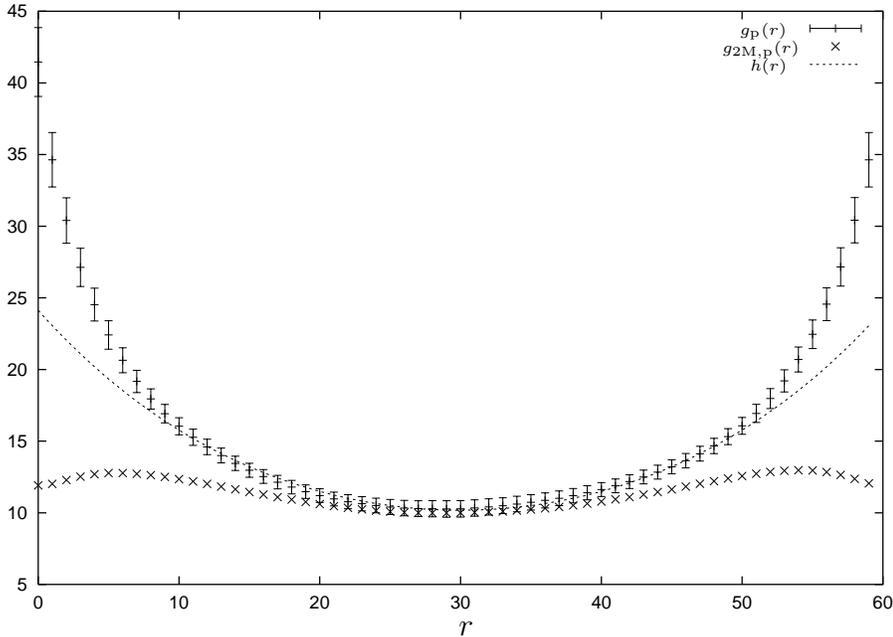,width=\linewidth}
\caption{\small \label{fullcorrf3} Projected correlation function
  $g_{\mathrm{p}}$ and two-meron part of the projected correlation function
  $g_{\mathrm{2M,p}}$ at $L = 60$ and $\beta=0.1$.}
\end{minipage}
\end{figure}

\begin{figure}
\noindent
\begin{minipage}[b]{.9\linewidth} 
\psfrag{fitfunf}{\tiny $\frac{3.138}{L}$}
\psfrag{massenpi}{\tiny $\qquad m_{\pi}$} 
\psfrag{1/L}{$1/L$}
\psfrag{m_pi}{$m_{\pi}$}
\centering\epsfig{file=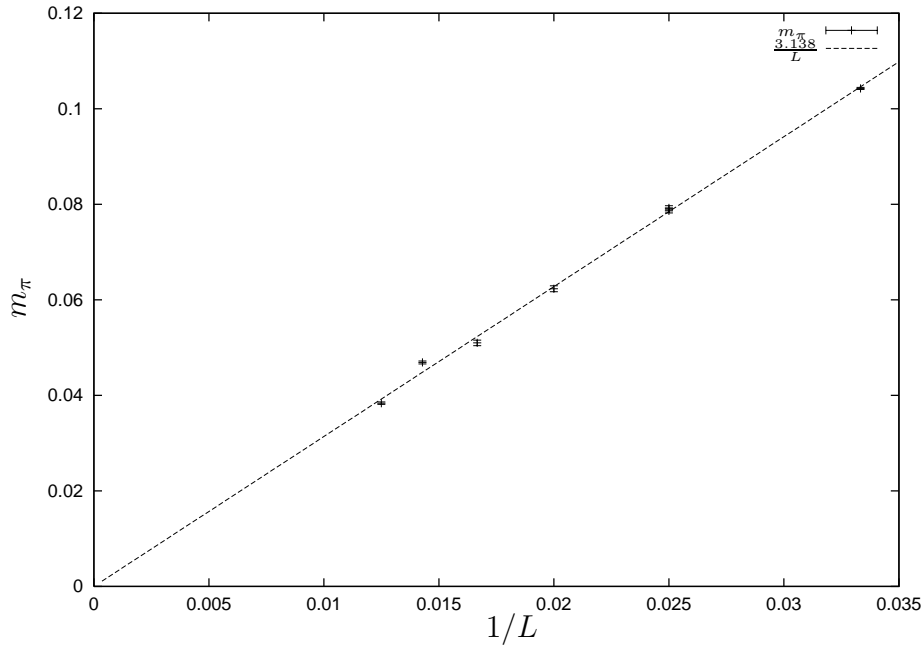,width=\linewidth}
\caption{\small \label{massen} Dependence of the mass $m_{\pi}$ on the
  lattice size $L$ at fixed $\beta=0.1$. The function $3.138/L$ is also
  plotted.}
\end{minipage}
\end{figure}

The meron cluster algorithm is  
used to investigate the large distance behavior of the correlation function
of two spins at $\theta=\pi$. 
We calculate the projected correlation function at $\theta=\pi$ 
\begin{equation}
g_{\mathrm{p}}(r)=
\left\langle (-1)^Q 
\left(\frac{1}{\sqrt{L_2}}\sum_{x_2=0}^{L_2-1}  e_{x_1,x_2}^{\parallel}\right)
\left(\frac{1}{\sqrt{L_2}}\sum_{y_2=0}^{L_2-1}  e_{y_1,y_2}^{\parallel}\right)
\right\rangle
\end{equation} 
(see figure \ref{fullcorrf3}) for various values of the lattice
 size $L=L_1=L_2$ at fixed $\beta=0.1$. 
The spins are summed over one
space direction (the 2-direction) in order to suppress the contributions
from excited states. This corresponds to a projection on the zero mode.
The data points are fitted with a hyperbolic cosine 
\begin{equation} \label{fitfunh}
h(r)=A\cosh\left(m_{\pi}\left(\frac{L}{2}-r\right)\right) .
\end{equation}
$A$ and $m_{\pi}$ are the fit parameters which must be determined.
The $L$~dependence of the masses $m_{\pi}$ obtained in this way is
 shown in figure
\ref{massen}. The $L$~dependence of $m_{\pi}$ seems to be approximately of 
the form
\begin{equation}
m_{\pi}=\frac{C}{L},\qquad C=3.138 .
\end{equation}
The value of the constant $C$ is obtained from a fit. 
These data suggest that in the infinite volume limit the mass $m_{\pi}$
vanishes and thus long
range correlations are present.
 The numerical data confirm the conjecture
that the correlation function at $\theta=\pi$
is long ranged (cf. section \ref{cfpi}). 
Also Haldane's conjecture about the behavior of quantum spin chains is
supported by the results  (cf. section \ref{intro}).
\section{Summary}
An improved estimator has been defined for the correlation
function of two spins in the two-dimensional O(3) model at $\theta=\pi$.
A meron cluster algorithm for the numerical calculation of the improved 
estimator has been developed. The numerical results indicate the occurrence of
long range correlations at $\theta=\pi$.
%
%
\section*{Acknowledgement}
The author thanks J. Cox, O. Jahn, F. Lenz and U.-J. Wiese for help and
discussions.
\end{document}